\documentclass[twocolumn,showpacs,preprintnumbers,amsmath,amssymb]{revtex4}
\usepackage{graphicx}
\usepackage{dcolumn}
\usepackage{bm}

\begin{document}

\title{Path Integral Method for DNA Denaturation}

\author{ Marco Zoli }
\affiliation{
Dipartimento di Fisica - Universit\'a di Camerino, I-62032, Italy. - marco.zoli@unicam.it}

\date{\today}

\begin{abstract}
The statistical physics of homogeneous DNA is investigated by the imaginary time path integral formalism. The base pair stretchings are described by an ensemble of paths selected through a macroscopic constraint, the fulfillement of the second law of thermodynamics. The number of paths contributing to the partition function strongly increases around and above a specific temperature $T^*_c$ whereas the fraction of unbound base pairs grows continuosly around and above $T^*_c$. The latter is identified with the denaturation temperature. Thus, the separation of the two complementary strands appears as a highly cooperative phenomenon displaying a smooth crossover versus $T$. The thermodynamical properties have been computed in a large temperature range by varying the size of the path ensemble at the lower bound of the range. No significant physical dependence on the system size has been envisaged. The entropy grows continuosly versus $T$ while the specific heat displays a remarkable peak at $T^*_c$.
The location of the peak versus $T$ varies with the stiffness of the anharmonic stacking interaction along the strand.
The presented results suggest that denaturation in homogeneous DNA has the features of a second order phase transition. The method accounts for the cooperative behavior of a very large number of degrees of freedom while the computation time is kept within a reasonable limit.

\end{abstract}


\maketitle

\section*{I. Introduction}

Path integral methods provide a powerful tool to investigate the properties of nonlinear dynamical systems with retarded interactions.  Hamiltonian models with non local electron-phonon couplings, used in polymer science \cite{io0}, have been studied by such methods.
In this paper the path integral approach \cite{fehi} is applied to analyse the temperature driven DNA denaturation which occurs when the double stranded molecule separates into two coils while, at lower $T$, only localized openings exist \cite{wart}. Modelling of DNA melting has been motivated since long  by the need to understand the transcription mechanism in which the double helix opens locally to allow a reading of the genetic code. In this regard, the Poland-Scheraga model \cite{poland} has been path-breaking and still makes the base of several investigations \cite{azbel2,peliti,stella,ambj,foge,livi}.  Despite extensive analytical and numerical work \cite{fisher,hanke,rud,rahi,whit,santos}, establishing character and nature of the melting phase transition, whether first or second order, remains a challenging task \cite{richard}.
A significant advance towards a comprehension of the DNA dynamics was made after Peyrard and Bishop (PB) introduced a 1D Hamiltonian model \cite{pey1} which recognized the role of nonlinearities in the molecule \cite{proh} and reduced its great complexity to two essential interactions: {\it i)} a nonlinear coupling between the two bases in a pair connected by hydrogen bonds, {\it ii)} a {\it harmonic} stacking potential between adjacent bases along the strand. While solutions based on the transfer integral method \cite{scala} predicted a smooth denaturation transition  both for homogeneous and heterogeneous DNA \cite{cule}, extensions of the PB model \cite{pey2,zhang} which include {\it anharmonic} stacking interactions showed that the denaturation can be sharp, occurs at lower $T$ than in the harmonic case and it is indeed a thermodynamic transition \cite{theo}.  The latter statement does not contraddict general theorems on the impossibility of phase transitions in 1D systems \cite{hove} as the PB Hamiltonian contains an on site potential depending on unbounded transverse stretchings. Moreover, in the thermodynamic limit, the energy of a domain wall between two states of the molecule is infinite and an ordered state is then possible \cite{landau}.

As the anharmonic PB model is assumed as fundamental tool, it is understood that the emphasis of the present investigation is purely on modelling the unbinding of the two strands versus temperature. This is done by capturing the essentials of the interactions consistently with the spirit of the theory of critical phenomena.

The PB Hamiltonian model is outlined in Section II focussing on those properties which make reliable to attack the problem by path integral techniques. Section III describes the method while the results concerning the character of the denaturation transition are discussed in Section IV. Some conclusions are drawn in Section V.

\section*{II. Hamiltonian Model}

To begin, I consider a fragment of homogeneous DNA whose strands are represented by a set of point masses corresponding to the nucleotides. The Hamiltonian  \cite{pey2} for a chain of N base pairs ({\it bps}) with mass $\mu$, reads:

\begin{eqnarray}
& & H =\, \sum_{n=1}^N \biggl[ {{\mu \dot{y}_{n}^2} \over {2}} +  V_S(y_n, y_{n-1}) + V_M(y_n) \biggr] \, \nonumber
\\
& & V_S(y_n, y_{n-1})=\, {K \over 2} g(y_n, y_{n-1})(y_n - y_{n-1})^2 \, \nonumber
\\
& & g(y_n, y_{n-1})=\,1 + \rho \exp\bigl[-\alpha(y_n + y_{n-1})\bigr]\, \nonumber
\\
& & V_M(y_n) =\, D \bigl(\exp(-a y_n) - 1 \bigr)^2 \, ,
\label{eq:1}
\end{eqnarray}

where $y_n$, the transverse stretching for the {\it n- bp}, measures the relative pair separation from the ground state position and continuously grows from $0$ (closed {\it bp}) to $\infty$ (open {\it bp} after denaturation occurs). As the hydrogen bond can be compressed with respect to the equilibrium, $y_n$ takes also negative values with a lower cutoff accounting for the repulsive interaction of the phosphate groups.
The longitudinal displacements are neglected in the model as their amplitudes are much smaller than the transverse stretchings \cite{wart,harvey}.  $D$ and $a$, setting depth and width of the Morse potential $V_M(y_n)$, are site independent for {\it homopolymer DNA}.
In the stacking potential $V_S(y_n, y_{n-1})$, the factor $g$ depends  on the sum of the stretchings of two adjacent bases. $K=\, \mu \nu^2$ with $\nu$ being the harmonic phonon frequency.
For $\rho = 0$ in Eq.~(\ref{eq:1}), the statistical mechanics can be worked out exactly through the transfer operator method \cite{scala} which, in continuum approximation and strong $K$ limit, defines a pseudo-Schr\"{o}dinger equation for a particle in a Morse potential \cite{pey3}. The latter has a discrete spectrum with localized eigenfunctions at $T < T_c$ being

\begin{eqnarray}
T_c =\,{{2 \sqrt{2 K D}} \over {K_B a}},
\label{eq:1a}
\end{eqnarray}

the temperature at which the lowest eigenvalue vanishes and even the ground state merges into the continuum. $K_B$ is the Boltzmann constant. For $T > T_c$ only delocalized states exist. Thus, even the simplest harmonic stacking model shows that a transition occurs and $T_c$ is identified as the temperature at which the strands separate.  Taking typical values for DNA parametrization \cite{campa,wart} such as $D=\,30 meV$, $a=\,4.2 {\AA}^{-1}$ and $K=\,60 meV {\AA}^{-2}$, one gets $T_c =\,331K$.
Anharmonicity ($\rho > 0$) has a special meaning in DNA stacking.
While the analytical form of $g$  may not be unique in order to reproduce the denaturation transition \cite{joy1}, the key property of the stacking potential is the following: whenever either one of the {\it bps} is stretched over a distance larger than $\alpha^{-1}$, the hydrogen bond breaks and the electronic distribution around the two pair mates is modified. Accordingly, the stacking coupling (along each strand) between neighboring bases in Eq.~(\ref{eq:1}) drops from $K(1 + \rho)$ to $K$. Then, also the next {\it bp} tends to open as both bases are less closely packed along their respective strands.
Here is the {\it cooperative} character which underlines the formation of a region with open {\it bps}, a bubble whose size increases with $T$. In heterogeneous DNA, also the length of the sequence may affect the probability to open a bubble which is quantified in polymer network theories by the {\it cooperativity parameter $\sigma$}: for sequences of intermediate length (up to $\sim 10^4 {\it bps}$), a small $\sigma$ ($\sim 10^{-4} - 10^{-5}$) generally suppresses bubble formation while large portions of the helix unbind close to the melting making the transition highly cooperative \cite{carlon,bar,krueg}. For long sequences, $\sigma$ is larger and small bubbles may form already below the transition which accordingly is expected to be less cooperative.

When the stacking decreases the vibrational mode between the two bases softens thus reducing its contribution to the free energy. Then, as confirmed by molecular dynamics simulations \cite{pey3}, the denaturation onset is signalled by a phonon mode softening which, in general \cite{krum,io1}, may point to the occurence of a phase transition.
For these reasons, anharmonic stacking models well account for cooperativity effects in the formation of open domains. Being the latter a collective phenomenon, computational methods have to include hundreds of base pairs in order to study the dynamics even of a molecule fragment and this requires considerable computational power. These facts suggest that models at intermediate scales are important and motivate the idea to investigate the DNA denaturation by path integral techniques.

\section*{III. Path Integral Method}

The path integral formalism  defines the time ($t$) evolution amplitude between two points, say "a" and "b", as a sum over {\it all} histories along which a system can evolve in going from "a" to "b". Each history is weighed by a phase factor, the exponential of the action associated to a given path \cite{feyn}.
At finite temperature, the thermal properties of the system can be derived by weighing the contributions by the particle paths $x(\tau)$ running along the imaginary time axis $\tau$, after an analytic continuation is performed: $\tau=\,it$.
Accordingly, in the statistical formalism, the Euclidean action $A[x(\tau)]$ replaces the mechanical canonical action and the partition function is an integral in the path phase space. Each path is weighed by a probability factor $exp(-A[x(\tau)])$. Only closed paths contribute to the statistical partition function, the integration being a trace integration \cite{kleinert}.
In the calculations, not {\it all} histories can be accounted for and, given the specific problem, one has to select the suitable class of paths which mainly contribute (are expected to contribute) to the physical properties.

For the model in Eq.~(\ref{eq:1}), the path integral description naturally follows by replacing the {\it bps} transverse stretching by a one dimensional path $x$ which is continuous function of $\tau$. Accordingly, the space interaction are mapped onto the time scale as

\begin{eqnarray}
& &y_n \rightarrow x(\tau), \, \, \, y_{n-1} \rightarrow x(\tau')\, , \, \tau' = \tau - \Delta \tau \, , \,
\label{eq:3}
\end{eqnarray}

with $\tau \in [0, \beta]$ and $\beta$ being the inverse temperature.
While only adjacent bases are taken in Eq.~(\ref{eq:1}), the stacking interactions range can be varied in the path integral model by tuning the retardation $\Delta \tau$ in Eq.~(\ref{eq:3}).
The discrete lattice nature of the Hamiltonian is maintained by the replacement in Eq.~(\ref{eq:3}). In fact, for any $\beta$, the computation is made convergent by taking a large but finite number $N_{\tau}$ of points in the $\tau$ range. The path $x(\tau)$ is further expanded in Fourier series with cutoff $M_F$

\begin{eqnarray}
x(\tau)=\, x_0 + \sum_{m=1}^{M_F}\Bigl[a_m \cos(\omega_m \tau) + b_m \sin(\omega_m \tau) \Bigr]
\label{eq:3a}
\end{eqnarray}

and $\omega_m =\, {{2 m \pi} / {\beta}}$. As $x(0)=\,x(\beta)$, periodic boundary conditions hold for the present method analogously to those imposed for the 1D finite chain described by Eq.~(\ref{eq:1}).  The heart of the matter lies in summing over a suitable set of paths such that the computation of the thermodynamical properties becomes sufficiently accurate.  As explained hereafter, the selection of the paths ensemble mainly contributing to the partition function is done on the base of physical constraints. By Eq.~(\ref{eq:3}), the index $n$ maps onto a single $\tau$ value and, by Eq.~(\ref{eq:3a}), for any $\tau$ an ensemble of path coefficients is taken in the computation.

The imaginary time partition function \cite{fehi} is given by

\begin{eqnarray}
& &Z=\,\oint \mathfrak{D}x(\tau)\exp\bigl[- A(x(\tau))\bigr]\, \nonumber
\\
& &A(x(\tau))=\,\int_0^\beta d\tau \Bigl[{\mu \over 2}\dot{x}(\tau)^2 + V(x(\tau)) \Bigr] \, \nonumber
\\
& &\oint \mathfrak{D}x(\tau)\equiv {1 \over {\sqrt{2}\lambda_\mu}}\int dx_0 \prod_{m=1}^{M_F}\Bigl({{m \pi} \over {\lambda_\mu}}\Bigr)^2 \int da_m \int db_m \, \, , \, \nonumber
\\
\label{eq:3b}
\end{eqnarray}

where $A(x(\tau))$ is the Euclidean action for a particle in the potential $V(x(\tau))$.
$\mathfrak{D}x(\tau)$ is the measure of integration which normalizes the free particle action

\begin{eqnarray}
\oint \mathfrak{D}x(\tau)\exp\Bigl[- \int_0^\beta d\tau {\mu \over 2}\dot{x}(\tau)^2  \Bigr] =\,1
\label{eq:6} \,
\end{eqnarray}

and $\oint$ denotes integration over closed particle trajectories consistently with periodic boundary conditions holding for paths in Eq.~(\ref{eq:3a}).

$\lambda_\mu$ is the thermal wavelength whose form depends on the model whether quantum or classical.  DNA denaturation occurs at temperatures for which a classical model applies. Then, the time derivative $\dot{y}_n$ (Eq.~(\ref{eq:1})) maps  onto the imaginary time derivative $\dot{x}(\tau)$ (Eq.~(\ref{eq:3b})),  the proper replacement being:  $d/dt \rightarrow ({\nu \beta}) d/d\tau$ hence, ${\lambda_\mu}=\,\sqrt{{\pi } / {\beta K}}$. Note that the pseudo-Schr\"{o}dinger equation  is also solved in a classical framework with $\hbar$ replaced by $({\nu \beta})^{-1}$.

Applying Eq.~(\ref{eq:3}) to the Hamiltonian in Eq.~(\ref{eq:1}), $Z$ in Eq.~(\ref{eq:3b}) transforms into

\begin{eqnarray}
& &Z=\,{1 \over {\sqrt{2}\lambda_\mu}}\int_{-U_0}^{U_0} dx_0 \prod_{m=1}^{M_F} \Biggl[
{1 \over {{\pi}}}\int_{-U}^U ds_m \int_{-U}^U dt_m \, \nonumber
\\
&\cdot& \exp(-s_m^2 - t_m^2) E(x_0, s_m, t_m)\Biggr] , \, \nonumber
\\
& & E(x_0, s_m, t_m)=\, {{\exp}}\Biggl(-\int_{0}^\beta d\tau V_{\Delta\tau}(x)\Biggr)   \, \nonumber
\\
& &V_{\Delta\tau}(x)=\, V_{M}(x) + V_{S}(x,x')\, \nonumber
\\
& &V_{M}(x)=\,D \bigl[\exp(-a x) - 1 \bigr]^2 \, \nonumber
\\
& &V_{S}(x,x')=\,{K \over 2} g[x,x'] \bigl(x - x'\bigr)^2 \, \nonumber
\\
& &g[x,x']=\,1 + \rho \exp\bigl[-\alpha\bigl(x + x'\bigr)\bigr] \, \nonumber
\\
& &x\equiv\, x(\tau) \, ; \, x' \equiv\, x(\tau')\, \nonumber
\\
& &s_m^2 \equiv \,{{m^2 \pi^3 a_m^2} \over {\lambda^2_\mu}};  \, \, t_m^2 \equiv \,{{m^2 \pi^3 b_m^2} \over {\lambda^2_\mu}} \, , \,
\label{eq:5}
\end{eqnarray}

where the measure has been written in a form suitable for computation by introducing the cutoffs $U_0$ and $U$ on the Fourier coefficients integrals \cite{io2}.
By inserting the expansion in Eq.~(\ref{eq:3a}), the l.h.s. of Eq.~(\ref{eq:6}) transforms into a product of Gau{\ss}ian integrals. Normalization of the latter gives the mathematical criterion to set $U_0$ and $U$ through the conditions:

\begin{eqnarray}
& &{1 \over {\sqrt{2}\lambda_\mu}}\int_{-U_0}^{U_0} dx_0 =\, 1 \, \nonumber
\\
& &{1 \over {\sqrt{\pi}}}\int_{-U}^U ds_m \exp(-s_m^2) =\, 1
\,
\label{eq:7}
\end{eqnarray}

While $U_0$ is set by the first in Eq.~(\ref{eq:7}), $U$ can be determined after a series expansion for the Gau{\ss}ian integral \cite{grad} is applied to the l.h.s. in the second of Eq.~(\ref{eq:7}). However, there is no {\it a priori} reason why the mathematical cutoffs had to fulfill also the physical requirements of the problem in Eq.~(\ref{eq:5}). In fact, as extensively discussed below, too negative Fourier coefficients would induce large negative path amplitudes and diverging $V_{M}$ which, in turn, would yield zero contribution to $Z$.  Such paths, although allowed by Eq.~(\ref{eq:7}), have therefore to be discarded as their inclusion in the computation would produce a decreasing entropy versus $T$. While this event cannot be accepted on general physical grounds, it also provides clue to solving consistently the cutoff problem:  imposing the second law of thermodynamics, I solve the path integral in Eq.~(\ref{eq:5}) selecting at any $T$ the suitable path ensemble for the transverse stretchings. Let's see in detail the technicalities of the method.

Besides the five {\it model potential} parameters, namely $D$, $a$, $K$, $\rho$, $\alpha$,   the path integral method contains some intrinsic input parameters. The latter are: {\bf a)} The cutoff $M_F$. {\bf b)} The retardation $\Delta \tau$.   {\bf c)} The number of points $N_\tau$ for the {$d\tau$}-integral in Eq.~(\ref{eq:5}). {\bf d)} The number of integration points over the Fourier coefficients $x_0$, $s_m$ and $t_m$.

{\bf a)} $M_F=\,1$ suffices to make the calculation convergent in the sense that $Z$ would not change by taking $M_F=\,2$. This result can be understood as follows: the inclusion of higher Fourier components leads to larger (absolute) values for the path amplitudes but $V_{M}$ {\it takes care of suppressing paths that leave the reasonable
boundaries}. While there is no reason to include paths whose fate is that of being suppressed, a proper ensemble of paths can be built by taking a sufficiently high number of integration points for the first Fourier component in a sufficiently broad range. This ensures that the programme samples a large portion of the phase space in which all the relevant path amplitudes are accounted for. Incidentally, the ($M_F=\,1$)-strategy also permits to save a great amount of computing time.

{\bf b)} This investigation is restricted to adjacent {\it bps} interactions along the strand. Accordingly, $\tau$ and $\tau'$ in Eq.~(\ref{eq:3}), are first neighbors in the discrete imaginary time lattice and the retardation is: $\Delta \tau =\,\beta / N_\tau$.

{\bf c)}  $N_\tau=\,100$ suffices to make the action numerically stable. This value also sets the number of {\it bps} as $N_\tau$ coincides with $N$ in Eq.~(\ref{eq:1}) due to the mapping in Eq.~(\ref{eq:3}). While
$N=\,100$ is a (minimal) reliable choice for a cooperative phenomenon such as denaturation, it turns out that larger $N_\tau$ ($N$) would not change anything in the computation of Eq.~(\ref{eq:5}). At this stage, the reader may wonder whether and how the path integral method can account for those finite {\it size} effects \cite{manghi,joy2}
which are considered relevant in studies of the DNA properties.  The fact is that, for the present method, $N$ is not the exclusive tuning parameter in order to set the system {\it size}, the latter being rather determined by the total number of paths allowed for the $N$-{\it bps}. Thus, although $N$ is kept constant, the system {\it size} may still increase/decrease by considering a larger/smaller path ensemble which eventually works as the {\it true scaling parameter} in the path integral method. Here, we see the deep meaning of the method underlied by the mapping in Eq.~(\ref{eq:3}). The $n-bp$ maps onto an effective ensemble of paths $N_{eff}$ taken at a specific $\tau$. The latter may be viewed in itself as an ensemble of base pairs. As this holds for any of the $N_\tau$ points, the {\it true scaling parameter} turns out to be $N_{\tau} \cdot N_{eff}$.

{\bf d) } Taking one Fourier component, the program evaluates at any $T$ the contribution to $Z$
by three integrals over Fourier coefficients in Eq.~(\ref{eq:5}).  If, for a given number of integration points,  the increasing entropy constraint is fulfilled then the {\it good paths} are included in the computation.  The number of {\it good paths} for a single $\tau$ is $N_{eff}$. The total number of {\it good paths} over which computations are carried out is $N_\tau \cdot N_{eff}$.

Clearly, the computing time depends on the number of paths required to stabilize the system. Many paths should contribute to the thermodynamical properties consistently with the previous observations regarding the cooperativity behavior in the double helix. Such number is expected to grow markedly around and above the transition. This would be a consistent hallmark for the reliability of the path integral approach to the DNA Hamiltonian.

\section*{IV. Discussion}

\begin{figure}
\includegraphics[height=7.0cm,angle=-90]{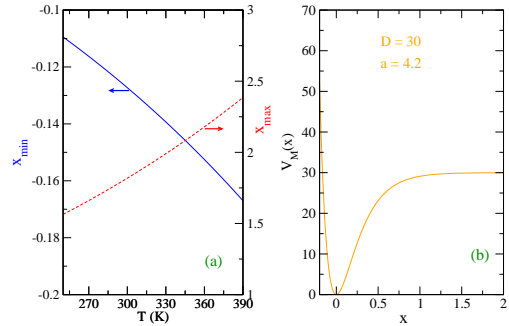}
\caption{\label{fig:1}(Color online) (a) Lower and upper cutoffs (in ${\AA}$) for the paths, representing the base pairs transverse stretchings, versus Temperature
; (b) Morse potential $V_M$ (in $meV$) versus path amplitude $x$.  $D$ is in $meV$ and $a$ in ${\AA}^{-1}$.}
\end{figure}

The issue of the maximum (absolute) path amplitudes is key to understand the double helix denaturation \cite{pey4} and it has to be discussed consistently with the available experimental data. As put forward above, the properties of the potential $V_{M}$, plotted in Fig.~\ref{fig:1}(b), suggest which path ensemble is physically meaningful.

Hydrogen bond stretchings are generally broken for lengths  $ \sim 2{\AA}$. For such separation, the potential is flat with a plateau value $D \sim 30meV$ and the force between the bases vanishes. Accordingly, larger amplitudes would yield no contribution. Then, the {\it bp} breaking energy $D$ is $\sim K_BT$ above room temperature. This justifies the parameter values used in Fig.~\ref{fig:1}(b). Largely positive paths would sample the flat portion of $V_M$ and add constant terms to the action which do not affect the free energy derivatives.

On the other side, steric hindrance processes prevent $x(\tau)$ from being too negative. Paths $x(\tau) \sim -0.14{\AA}$ also experience a value $D \sim 30meV$ while larger negative paths encounter the hard core repulsive side of $V_M$. Here the latter grows exponentially thus yielding a large action which, in turn, makes a vanishing contribution to $Z$.

These are the intuitive reasons to set a finite range which inspire the computational method described in the previous Section. Moreover, it is physically plausible that such range had to be modulated by temperature effects allowing for larger path amplitudes at higher $T$. Saying $T^*_c$ the denaturation temperature, the path displacements may reasonably vary in the range $x(\tau) \in [x_{min}, x_{max}]$  with $x_{min} \sim -0.14 {\AA}$ and $x_{max} \sim 2{\AA}$ around the crossover.
The functions: $\, \, x_{min}=\, -0.14 \exp[ (T-T^*_c)/T^*_c]\, \,$ and $\, \,x_{max}=\, 2 \exp[ (T-T^*_c)/T^*_c]\, \,$, provide suitable (although not unique) expressions to account for linear temperature dependence in a window around $T^*_c$ whereas larger deviations occur outside that window.  $x_{min}$ and $x_{max}$ are plotted in Fig.~\ref{fig:1}(a)  taking for $T^*_c$ the value given by Eq.~(\ref{eq:1a}).
However, this assumption  does not imply that the melting occurs precisely at $T_c$. In fact the true  denaturation temperature (found below) does not coincide with the above given $T_c$ which stems from an analytic harmonic model.
Moreover, the specific choice for upper and lower bounds has not to imply in principle that a transition exists nor it has to force the transition to take place at $T^*_c$.
Certainly, if the denaturation occurs, a significant fraction of {\it bps} is expected to be $ \sim 2 {\AA}$. This feature will be hereafter investigated.

\begin{figure}
\includegraphics[height=7.0cm,angle=-90]{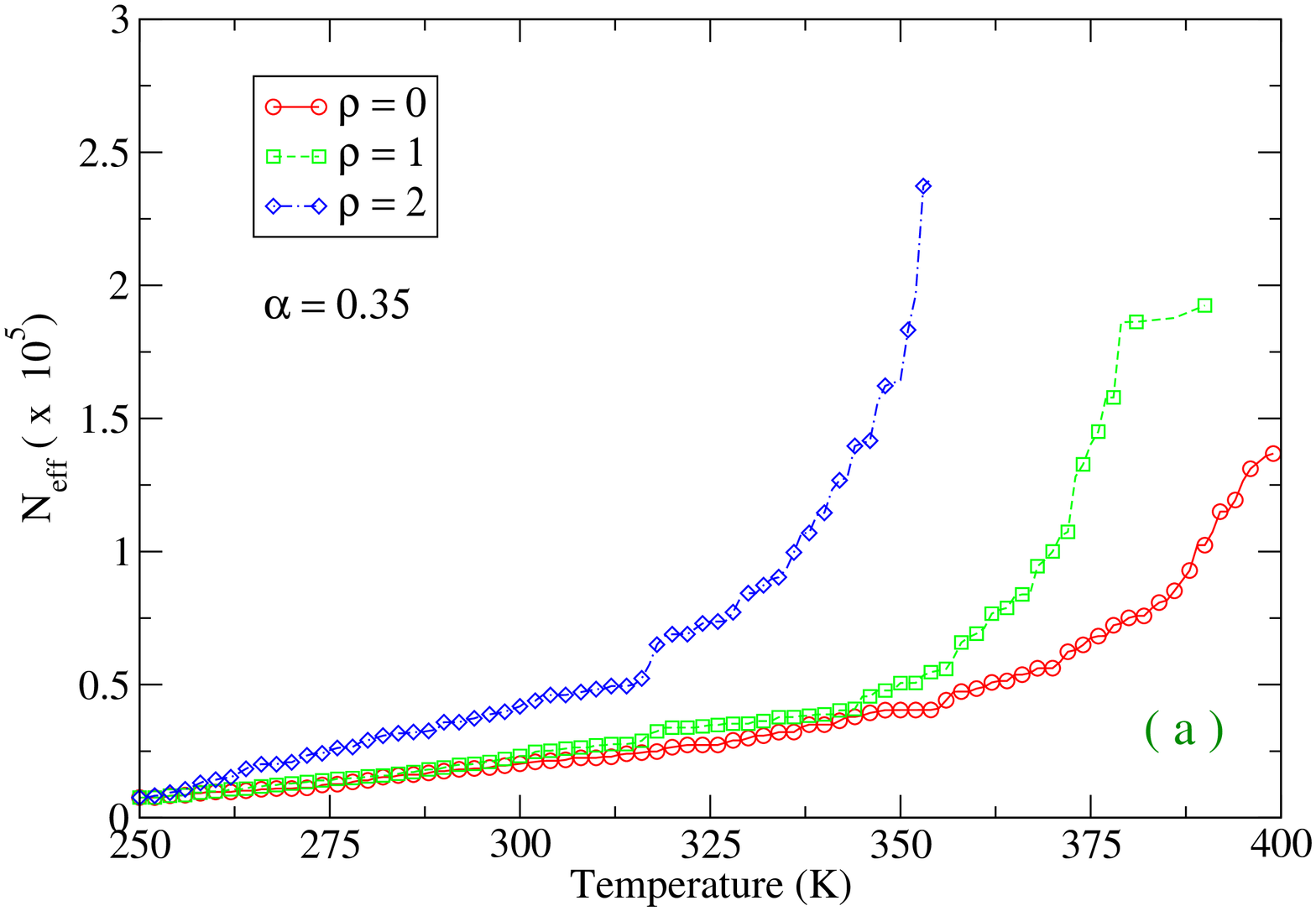}
\includegraphics[height=7.0cm,angle=-90]{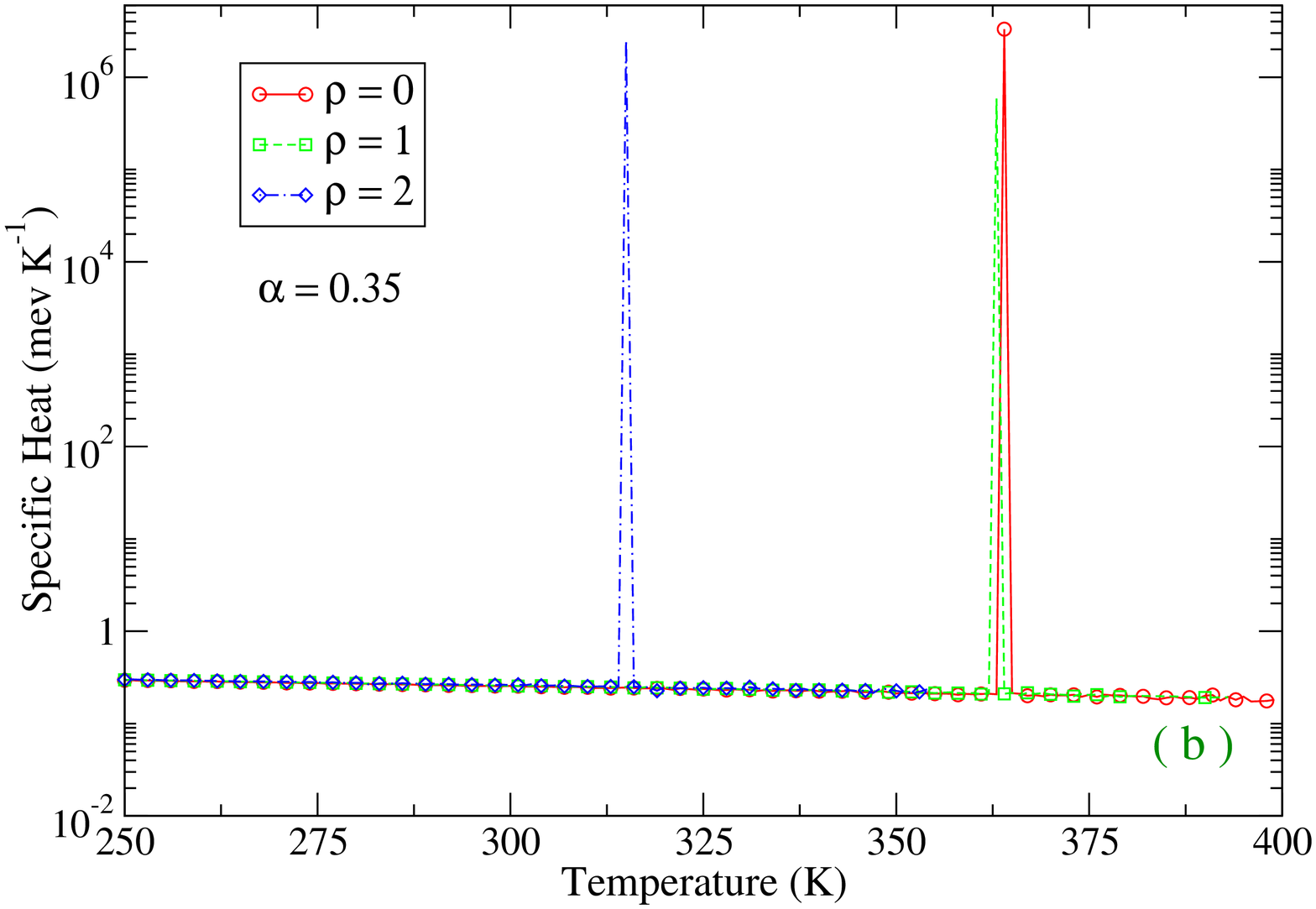}
\caption{\label{fig:2}(Color online) (a) Effective Number of Paths contributing to the partition function for three stacking parameters. $\alpha$ in ${\AA}^{-1}$. (b) Specific heat computed via Eq.~(\ref{eq:5}). }
\end{figure}

Thus, after setting a temperature value, the program searches for those paths $\{ x(\tau),x(\tau') \}$ which lie inside the range $[x_{min}, x_{max}]$ for any
$\tau \in [0, \beta]$ and adds their contribution to $Z$. Such a check is done for any set of Fourier coefficients which in fact defines a specific path. Had the entropy not to have a positive derivative, the computation should be re-run for a new {\it number of integration points} which, in turn, leads to re-build the ensemble of Fourier coefficients and to repeat the procedure to filter the suitable paths. It is seen numerically that the restriction due to $x_{max}$ can be lifted without affecting the physical properties whereas the lower bound $x_{min}$ is required in the computational method.

As a final output $N_{eff}$ is obtained versus $T$ as shown in Fig.~\ref{fig:2}(a) for the same values $D$, $a$, $K$ as in Figs.~\ref{fig:1} and for three stacking parameters $\rho$ with $\alpha=\,0.35 {\AA}^{-1}$ \cite{campa}.  While all plots begin with $N_{eff} \simeq 8000$ at $T=\,250K$, their slopes display a remarkable dependence on $\rho$ and, as a general feature, $N_{eff}$ is higher for stronger anharmonicities.
In particular, for the case $\rho=\,2$, $N_{eff}$ increases at $T \sim 315K$. This feature is mirrored in the plot (Fig.~\ref{fig:2}(b)) of the specific heat which is a significant hallmark for denaturation as it is proportional to the differential melting curve \cite{coluzzi}. For $\rho=\,2$, a peak is found at the same $T$ and in general, the specific heat displays a sharp increase when the degree of cooperativity (measured by $N_{eff}$) markedly grows. The peak shifts towards lower $T$ by increasing $\rho$. This finding, shared by previous studies \cite{pey2}, is physically meaningful as systems with larger anharmonic couplings  should have a higher number of pairs which break at lower $T$.

\subsection*{A. Phase Transition}

While the results displayed so far witness by themselves reliability and consistency of the method, they also induce to face the unresolved issue regarding the order of the denaturation transition. Two questions need to be addressed:

{\bf 1)} how does the entropy behave?

{\bf 2)} To which extent do the thermodynamical properties depend on the number of paths included in the calculation? Or, how does the macroscopic system scales versus $N_{eff}$?

Taking the intermediate plot,  $\rho=\,1$ in Fig.~\ref{fig:2}, I have thus progressively increased $N_{eff}$ over the value $N_{eff} =\, 9920$ at $T=\,260K$ in Fig.~\ref{fig:2}(a). While the $T$ axis in the latter starts from $250K$, the $10K$ shift upwards allows one to distinguish among the three plots. $T=\,260K$ is then the lower bound in the $T$ window considered hereafter. Once the starting $N_{eff}$ is set at the lower bound, the thermodynamical properties are evaluated at higher $T$ according to the procedure detailed above.

\begin{figure}
\includegraphics[height=7.0cm,angle=-90]{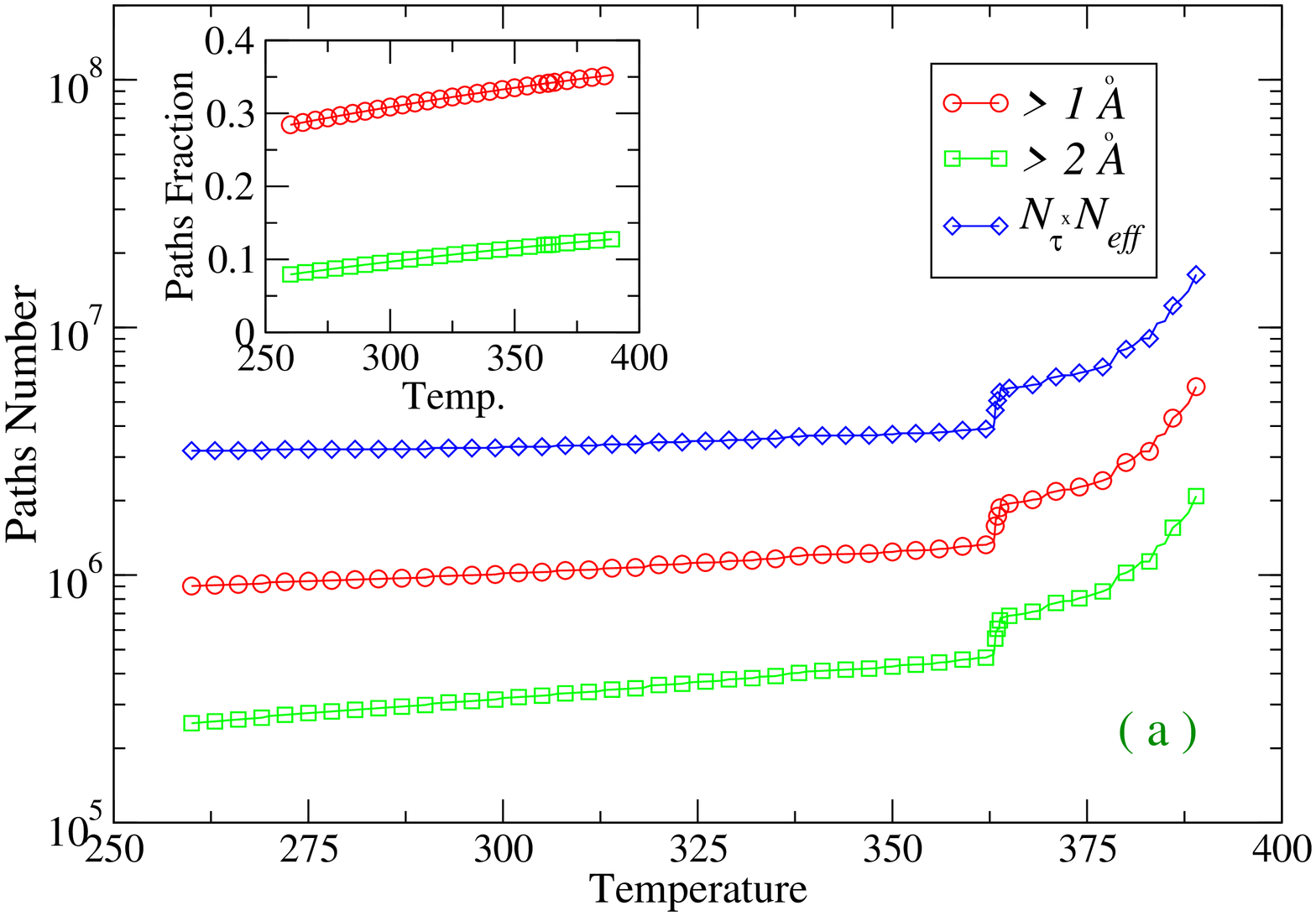}
\includegraphics[height=7.0cm,angle=-90]{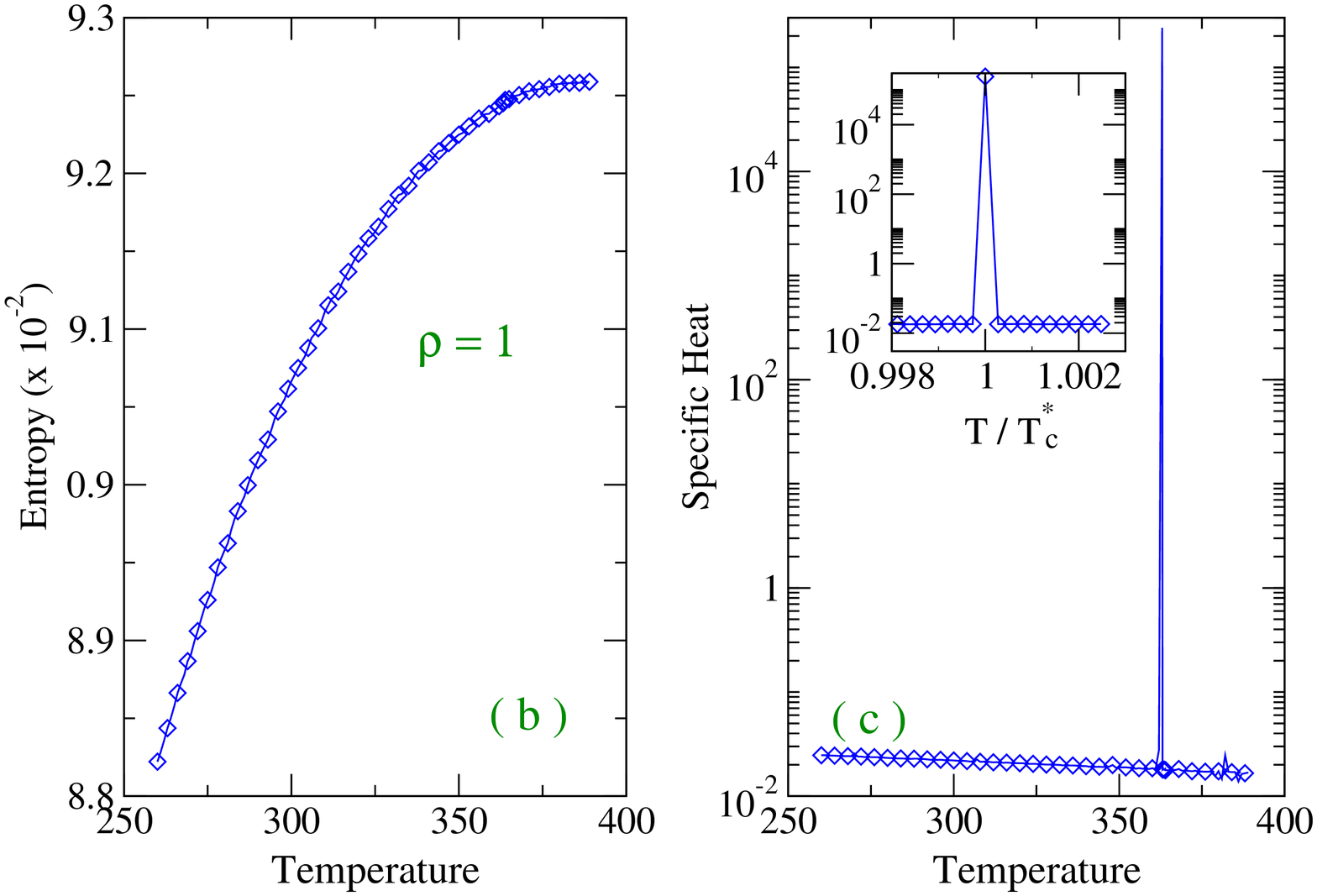}
\caption{\label{fig:3}(Color online) Anharmonic stacking case $\rho=\,1$. $N_{eff}=\,31792$ at $T=\,260K$ is assumed. $T$ axis starts with a $10K$ shift with respect to Fig.~\ref{fig:2} in order to distinguish among the three plots corresponding to different $\rho$ values. (a) Paths larger than $1{\AA}$, $2{\AA}$ and {\it Total Number} of paths ($N_\tau \cdot N_{eff}$) are plotted. Inset: ratios of paths whose amplitude is larger than $1{\AA}$ and $2{\AA}$ respectively, over $N_\tau \cdot N_{eff}$.  (b) Entropy versus Temperature.  (c) Specific Heat versus Temperature. Inset: Magnification of the Specific Heat Peak with a temperature resolution of $0.1K$.}
\end{figure}

In Fig.~\ref{fig:3}, the calculation begins with $N_{eff}=\,31792$, a value larger by a factor three than in Fig.~\ref{fig:2}.   Fig.~\ref{fig:3}(a) plots:  {\it i)} the total number of paths versus $T$ starting with $\sim 3.2 \cdot 10^6$ at $T=\,260K$; {\it ii)} the number of paths whose amplitude is larger than $1{\AA}$; {\it iii)} the number of paths whose amplitude is larger than $2{\AA}$. All curves increase versus $T$ and show a kink at $T^*_c$. This feature is more pronounced and evident than in Fig.~\ref{fig:2} confirming that denaturation is indeed a highly cooperative phenomenon. While paths $> 1{\AA}$ already sample the plateau of $V_M$ where the {\it bps} unbind, paths $> 2{\AA}$ certainly belong to broken pairs. Accordingly, the latter become more numerous around and above $T^*_c$. Thus, approaching the transition, the system incorporates a larger number of paths with an increasing fraction of broad path amplitudes. As such fractions are considered  reliable indicators of the order of the denaturation transition, quantitative estimates are greatly relevant.  As the inset makes evident, the fractions of paths $> 1{\AA}$ and $> 2{\AA}$ respectively (normalized over $N_\tau \cdot N_{eff}$), grow continuosly versus $T$. Around $T^*_c$, one third of all paths are larger than $1{\AA}$: it means that either they are broken or about to break.   We see no step-like increase at $T^*_c$ and this behavior is fully consistent with the continuous entropy growth shown in Fig.~\ref{fig:3}(b). The slight irregularity appearing in the entropy plot at $T^*_c$ is responsible for the remarkable peak in the specific heat plotted in Fig.~\ref{fig:3}(c). The peak remains pinned at $T^*_c=\,363K$ as in Fig.~\ref{fig:2}(b).  The inset shows an enlargement with a temperature resolution which appreciates $0.1K$. Finer partitions may be taken further increasing computing time.

\begin{figure}
\includegraphics[height=7.0cm,angle=-90]{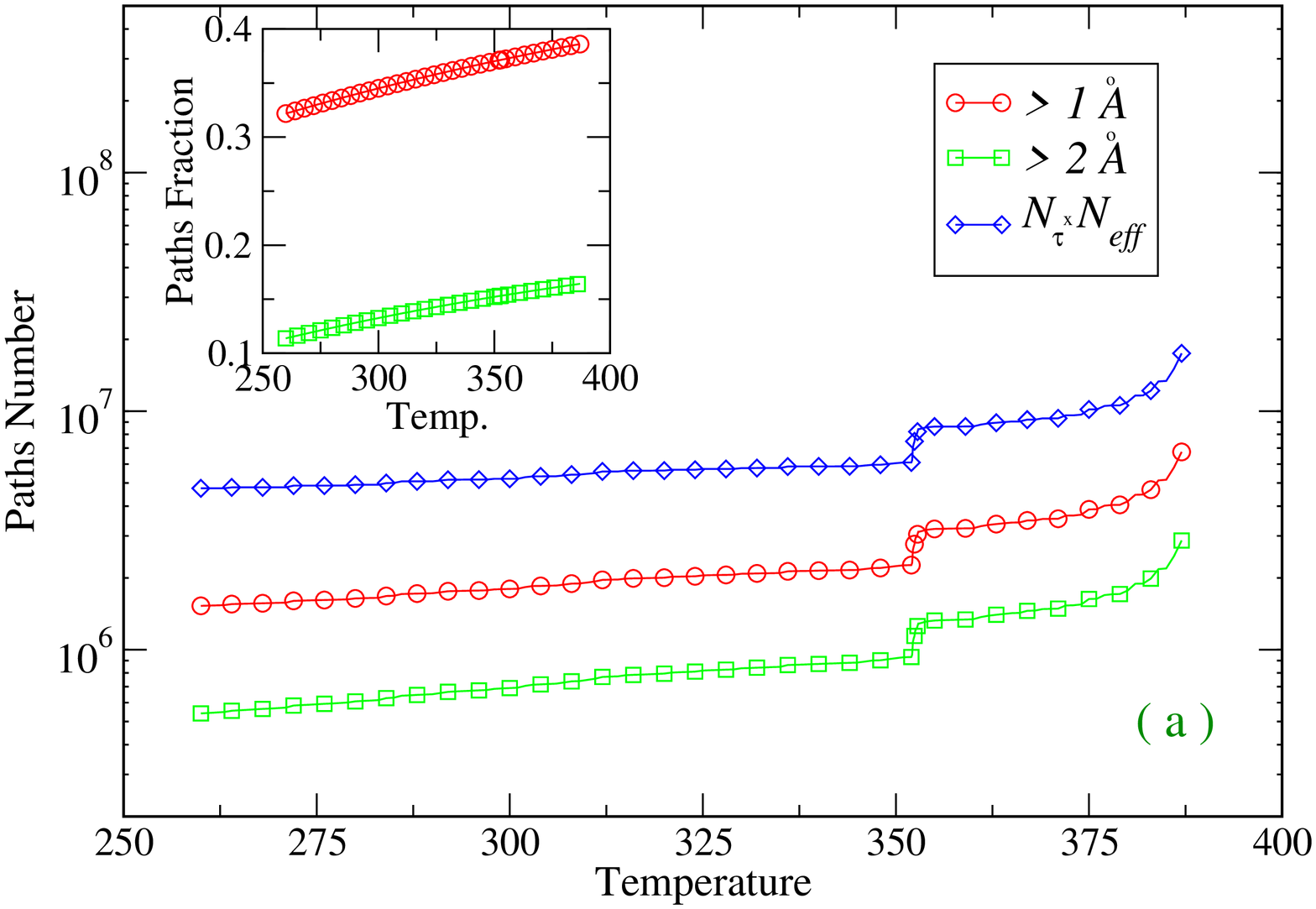}
\includegraphics[height=7.0cm,angle=-90]{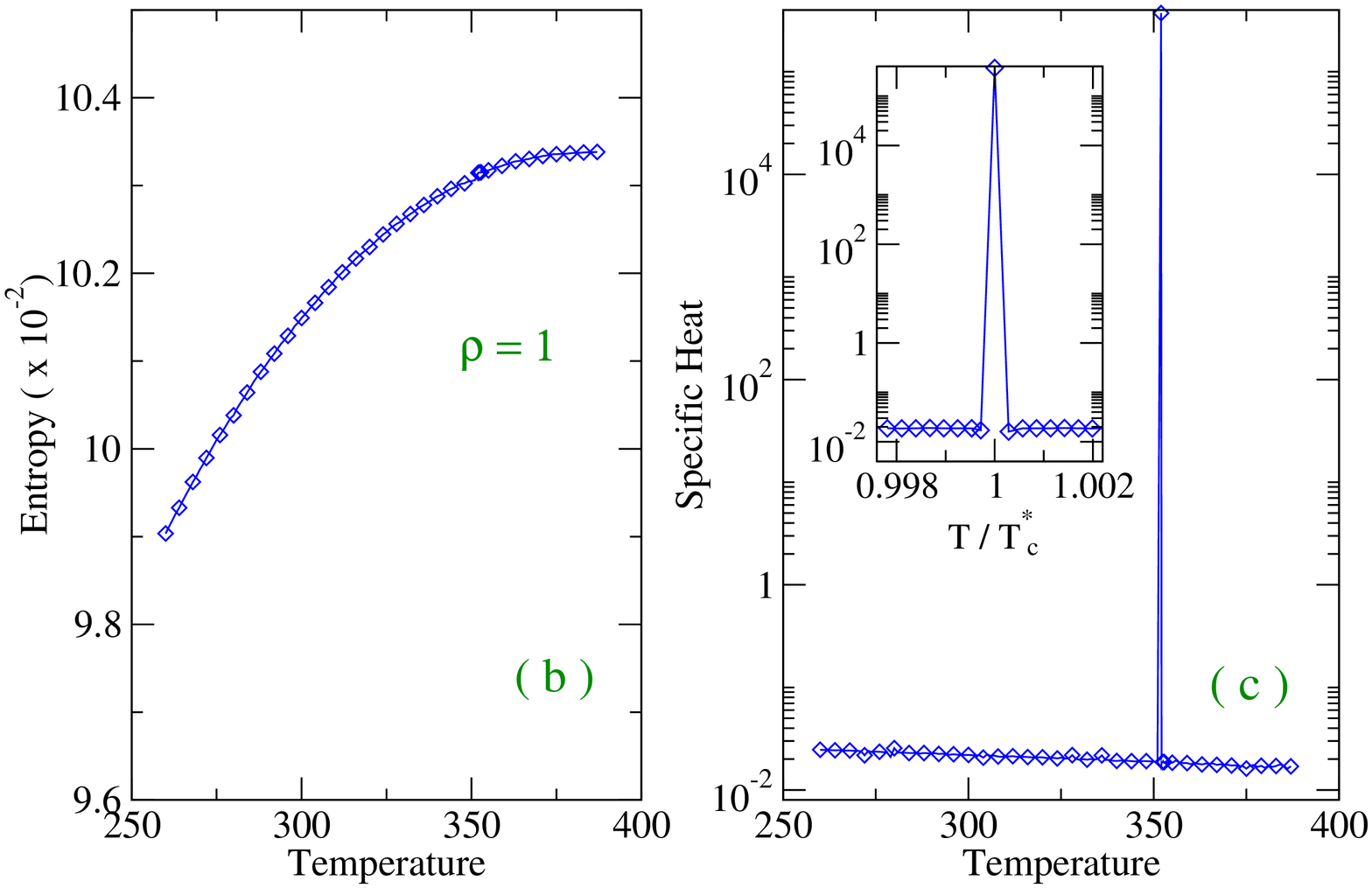}
\caption{\label{fig:4}(Color online) As in Fig.~\ref{fig:3} but with $N_{eff}=\,47494$ at $T=\,260K$. }
\end{figure}

The results presented so far point to the occurence of a smooth crossover at $T^*_c$. While evidence is emerging for classyfing the transition as of second order, the careful reader may argue that the aforementioned question {\bf 2)} is not yet fully answered. I have then further increased the starting $N_{eff}$: all the obtained curves look like those presented in Fig.~\ref{fig:3} with, at most, shifts of a few degrees in the location of $T^*_c$. Fig.~\ref{fig:4} shows the results obtained for $N_{eff} =\, 47494$ at $T=\,260K$: in this case $T^*_c=\,352K$  which makes the largest observed shift with respect to Figs.~\ref{fig:2} and ~\ref{fig:3}. Note that $\sim 8 \cdot 10^6$ paths participate to the computation around $T^*_c$, with $37 \%$ being larger than $1{\AA}$. Remarkably, $\sim 18 \cdot 10^6$ paths are accounted for at the largest $T$ here considered.

Some features shown in the plots and drawn from the whole numerical work are in order:

{\bf *)} The entropy is always a continuosly increasing function of $T$ whatever the starting value for $N_{eff}$ may be.

{\bf *)} At $T^*_c$, the entropy always displays a slightly more pronounced enhancement although there is no finite melting entropy which would point to a first order phase transition \cite{theo}. The small entropy kink is instead responsible for the peak in the specific heat.  The crossover is smooth, the phase transition appears to be of second order.

{\bf *)} The entropy values are larger by including more paths in the computation.

{\bf *)} Above $T^*_c$, the entropy goes on growing but with smaller gradient consistently with the fact that, as the strands separation proceeds, the probability to add entropic effects is reduced.

{\bf *)} The location of $T^*_c$ slightly varies with increasing $N_{eff}$ but the changes are not significant oscillating in the range $T^*_c \simeq [350 - 365K]$ for $\rho=\,1$.

While different choices for the stiffness parameters in the PB Hamiltonian are not expected to change these qualitative physical features, it may be instead worth investigating by path integrals also Hamiltonian models with different stacking potentials  both for homogeneous and heterogeneous DNA \cite{joy3}.

\section*{V. Conclusion}

The homogeneous DNA denaturation has been studied in the path integral formalism by mapping the essential interactions of the Peyrard-Bishop Hamiltonian onto the imaginary time scale. While the nonlinearities in the stacking interactions are recognized as a key feature of the Hamiltonian, the path integral method permits to deal with the highly cooperative character of the denaturation by incorporating a very high number of degrees of freedom. I have thought of the transverse stretchings for the base pairs as time dependent paths: in the imaginary time formalism, this amounts to incorporate the temperature effects in the base pairs displacements. The path amplitudes are taken consistently with the shape of the Morse potential modelling the hydrogen bonds. Accordingly, I have set up a computational method to monitor, versus temperature, the paths ensemble which mainly contributes to the partition function, hence to the macroscopic equilibrium properties of the system. A temperature window has been considered in which denaturation is expected to occur. Given the size of the paths ensemble at the lower bound of such window, the second law of thermodynamics has been invoked to build the ensemble at any larger $T$. The method works self-consistently selecting  the minimum number of paths required (at any $T$) to pursue the goal, that is the growing entropy constraint. There is no need to add further paths to the partition function once such goal is achieved: I feel that a {\it minimum effort principle} should hold. Instead, the system physical properties may depend on the initial ensemble size and care has been taken to unravel this issue.
Thus, the partition function and its derivatives have been computed by varying the boundary condition but no significant variation in the physical behavior has been observed. It is understood that the initial size should be large enough to allow for the appearance of cooperative effects. Remarkably, the ensemble size strongly grows around and above the denaturation temperature.  The denaturation of the complementary strands always shows up as a smooth crossover whose location along the $T$-axis may depend on the stiffness of the stacking interaction. Focussing on the system with moderate anharmonic stacking, it has been found that: {\it i)} the entropy grows continuosly versus $T$, {\it ii)} the specific heat displays a peak precisely at the same temperature for which the path ensemble size sharply grows.

These results suggest that the denaturation in homogeneous DNA is a highly cooperative phenomenon with the hallmarks of a second order phase transition. Consistently, the fraction of paths sampling the plateau of the Morse potential grows continuosly also around the denaturation temperature. No step-like increase has been observed in the paths fractions which describe the unbinding of the base pairs.  Eventually, I wish to emphasize that the thermodynamical constraint applied on the system does not force {\it a priori} the entropy to grow continuosly. While the system autonomously selects at any $T$ the total number of paths to fulfill such constraint, both a step discontinuity and a continuous behavior may in principle appear. The obtained plots are therefore not an artifact of the method.

After testing the feasibility of the path integral approach to a large size system as homogeneous DNA, I feel that the presented formalism may be further improved/extended to account for bubble formations in heterogeneous models with inhomogeneous sequences.

\end{document}